\documentclass[manuscript,screen,nonacm]{acmart} 
\setcopyright{cc} 
\acmConference[]{} 

\newcommand{\workshopname}{GenAICHI: CHI 2025 Workshop on Generative AI and HCI}
\newcommand\extrafootertext[1]{
    \bgroup
    \renewcommand\thefootnote{\fnsymbol{footnote}}%
    \renewcommand\thempfootnote{\fnsymbol{mpfootnote}}%
    \footnotetext[0]{#1}%
    \egroup
}

\AtBeginDocument{ 
    \fancypagestyle{firstpagestyle}{
        \fancyhf{}
        \fancyfoot[L]{\sffamily\footnotesize \workshopname}%
        \fancyfoot[C]{\sffamily\footnotesize \thepage}
    }
    \fancyhf{}
    \fancyhead[L]{\sffamily\footnotesize\shorttitle}
    \fancyhead[R]{\sffamily\footnotesize\shortauthors}
    \fancyfoot[L]{\sffamily\footnotesize\workshopname}%
    \fancyfoot[C]{\sffamily\footnotesize\thepage}
}

\begin{document}

\title{BCause: Human-AI collaboration to improve hybrid mapping and ideation in argumentation-grounded deliberation}

\author{Lucas Anastasiou}
\email{lucas.anastasiou@open.ac.uk}
\orcid{0000-0002-1587-5104}
\affiliation{%
  \institution{The Open University}
  \streetaddress{Walton Hall}
  \city{Milton Keynes}
  \country{United Kingdom}
  \postcode{MK7 6AA}
}

\author{Anna De Liddo}
\email{anna.deliddo@open.ac.uk}
\orcid{0000-0003-0301-1154}
\affiliation{%
  \institution{The Open University}
  \streetaddress{Walton Hall}
  \city{Milton Keynes}
  \country{United Kingdom}
  \postcode{MK7 6AA}
}

\begin{abstract}
Public deliberation, as in open discussion of issues of public concern, often suffers from scattered and shallow discourse, poor sensemaking, and a disconnect from actionable policy outcomes. This paper introduces BCause, a discussion system leveraging generative AI and human-machine collaboration to transform unstructured dialogue around public issues (such as urban living, policy changes, and current socio-economic transformations) into structured, actionable democratic processes. We present three innovations: (i) importing and transforming unstructured transcripts into argumentative discussions, (ii) geo-deliberated problem-sensing via a Telegram bot for local issue reporting, and (iii) smart reporting with customizable widgets (e.g., summaries, topic modelling, policy recommendations, clustered arguments). The system's human-AI partnership preserves critical human participation to ensure ethical oversight, contextual relevance, and creative synthesis.
\end{abstract}


\begin{teaserfigure}
    \includegraphics[width=\textwidth]{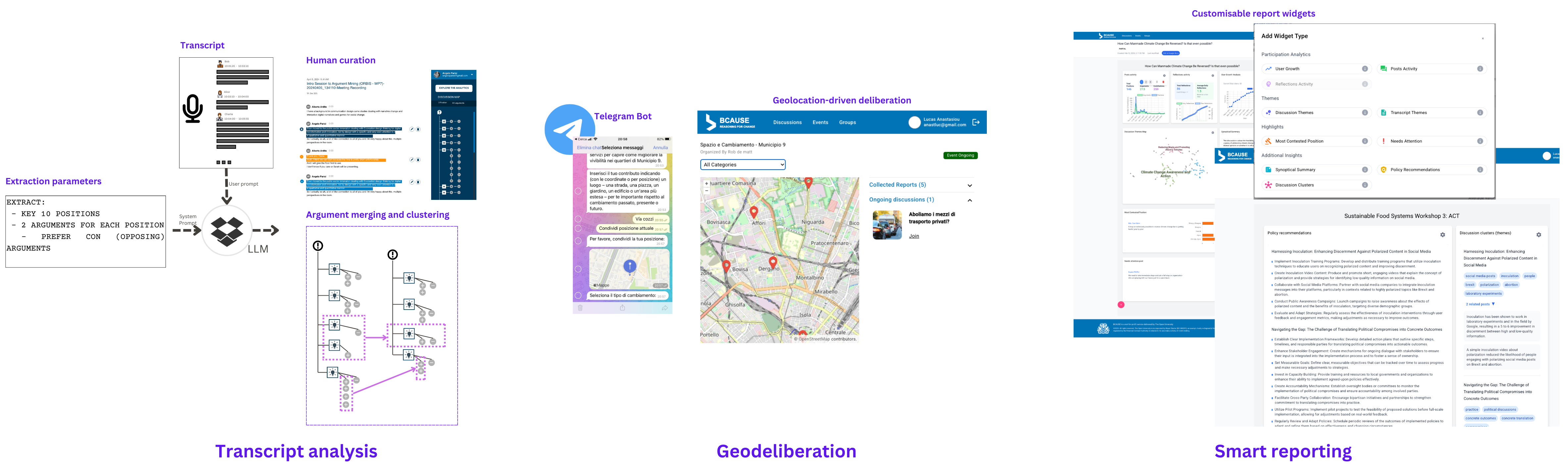}
    \caption{Overview of BCause's three main AI-enhancements: (left) transcript analysis and transformation into argumentative format, (centre) Geo-deliberation interface location-based issue reporting, and (right) Smart reporting dashboard with customizable widgets}
    \Description{figure description}
\end{teaserfigure}
\maketitle

\section{Introduction}

The landscape of public discourse faces unprecedented challenges in the digital age. Despite the proliferation of online discussion platforms, there exists a critical disconnect between public deliberation and policy formation processes \cite{hartz2014unfulfilled}. 
As \cite{magu2024understanding} observe, 
controversial online discussions about issues of greater public concern (such as gun policy, human rights violations, political extremism, etc) remain fragmented across disparate social media platforms, lacking coherent mechanisms for sensemaking, and rarely achieving meaningful influence on policy decisions. This fragmentation and ineffectiveness of online political discourse presents a significant impediment to democratic participation and informed decision-making in modern society \cite{freelon2015discourse}.

To address this challenge, deliberative democracy \cite{bachtiger2018oxford} relies on collective sensing mechanisms as a crucial methodology for individuals and community groups to take an active role in gathering and recording information about their lived environment, which can range from personal notes and observations to large-scale data collection from numerous participants, providing insights into broader sentiments, trends and patterns, for instance, across a city \cite{cai2009spatial,goldman2009participatory}. This process is typically operationalised through the systematic crowdsourcing of individual participation in matters of civic significance \cite{brabham2013using}.

The integration and extension of geospatial components to the discussion of public issues has been known in HCI as geo-deliberation~\cite{carroll2013wild}, and provides a framework to enhance this process by anchoring public discourse to specific geographical data and contexts, allowing for spatially-informed collective decision-making \cite{jankowski2016geo}.

Leveraging recent advancements in AI and in generative AI particular~\cite{sengar2024generative}, we present enhancements in BCause\footnote{https://bcause.app}\cite{anastasiou2023bcause}, an argumentative discussion system designed to transform unstructured online political dialogue into structured geo-deliberation while establishing concrete pathways to policy influence. These enhancements in BCause respond directly to the growing need for digital platforms to effectively bridge the gap between public discourse and policy formation, while maintaining the accessibility and engagement that characterizes successful hybrid (physical and online) deliberation spaces.

\section{BCause System Overview}

BCause is a platform for discussion-enhanced collective intelligence that structures online deliberation through a layered argumentative approach~\footnote{https://bcause.kmi.open.ac.uk/}. At its core, it organises discussions around clearly defined positions, where participants can contribute supporting (pro) and opposing (con) arguments - a lightweight Issue-based Information System schema \cite{kunz1970issues}. These argumentative structures are visualised through a time-ordered argumentative tree, an innovative user interface designed to achieve the best of both linear and graphical argumentative interfaces. To aid collective understanding, BCause generates synoptical summaries, visualises the argument tree and identifies ``Sensemaking nuggets'' — insightful contributions that help advance the deliberation.

\section{GenAI in BCause}

\subsection{Import and Transformation of Unstructured Dialogue}

\begin{figure}[htbp]
    \centering
    \begin{minipage}{0.35\textwidth}
        \centering
        \includegraphics[width=\textwidth]{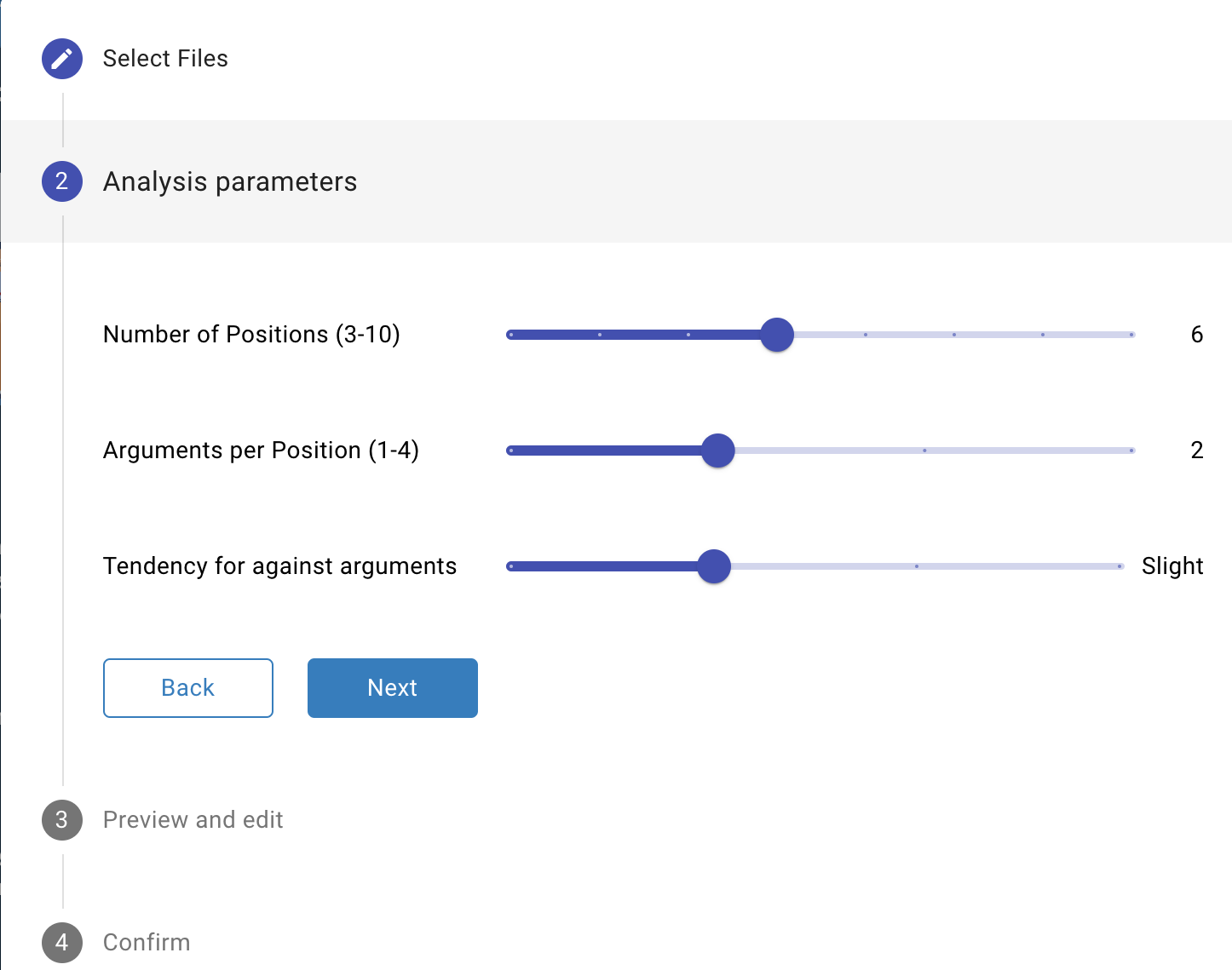}
    \end{minipage}
    \hfill
    \begin{minipage}{0.35\textwidth}
        \centering
        \includegraphics[width=\textwidth]{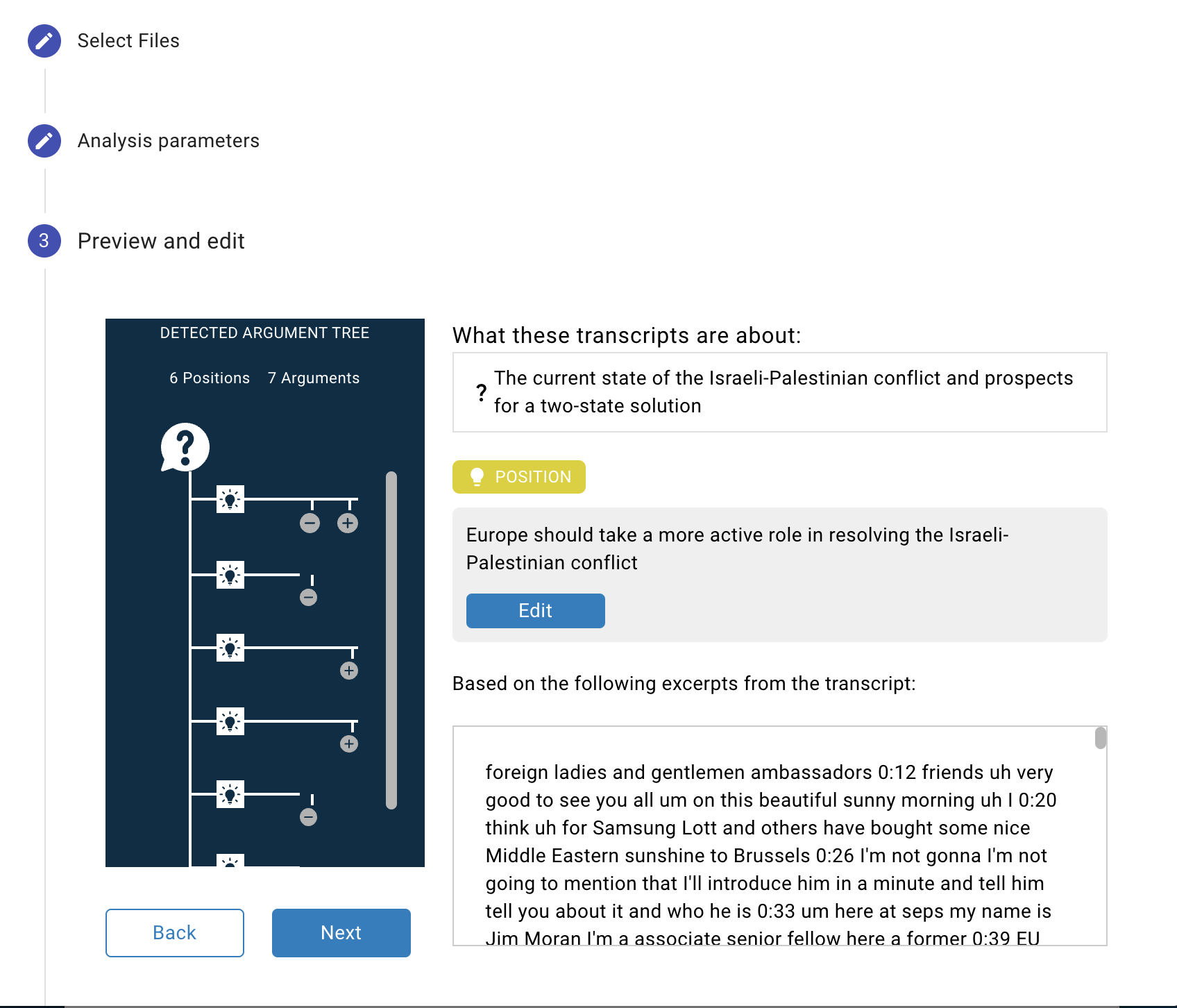}
    \end{minipage}
    \caption{Transcript analysis intermediate steps: (left) configurable parameters for argument extraction, including number of positions per issue, number of arguments per position and argument balance settings, (right) preview panel of the resulting structure}
    \label{fig:two-imports}
\end{figure}

To help bridge asynchronous modes of discussion (such as online discussion forums) with synchronous/live deliberation (both face-to-face events and online virtual meetings), BCause employs an automatic transcript processing system that can ingest recordings of live events and systematically organise the content into a structured argumentative framework. The system categorizes discussion elements into the key components of the IBIS schema: \textit{Issues} - High-level topics under deliberation (e.g., urban housing policy), \textit{Positions} - Specific stances on the issues (e.g., support for affordable housing mandates) and \textit{Arguments} - Evidence and reasoning provided to support or oppose positions (e.g., economic impact analysis).

The platform's innovation lies in its ability to automatically detect these argumentative components from natural discussion through a supervised machine learning approach. This automated structuring serves as a bridge between synchronous and asynchronous modes of deliberation - live discussion transcripts can be processed into structured arguments that seamlessly integrate into ongoing online debates.
Crucially, BCause implements a ``human-in-the-loop'' approach where AI-generated argument structures are presented as initial proposals rather than final determinations. Discussion moderators maintain oversight and can refine or correct the system's interpretation of argumentative relationships. The platform provides adjustable parameters for controlling aspects like: (i) Number of positions extracted per issue (from 3 to 10), (ii) Balance between supporting and opposing arguments (adjusting the bias to select opposing or supporting arguments) and (iii) number of arguments per position (min 1 to 4 max).

\subsection{Geo-Deliberated Problem Sensing via Telegram Bot}

BCause implements a geo-deliberated issue reporting mechanism through a Telegram bot interface, currently deployed in Milan's Municipio 9 district, part of the ORBIS\footnote{https://orbis-project.eu/} project's ``Space and Change'' initiative. The system architecture consists of three main layers: 1. \textit{Data Collection}: Citizens use a Telegram bot to submit geolocated reports including text descriptions (or audio messages) and optional images of local issues, 2. \textit{LLM-based classification} to categorize reported problems into municipal service categories, with the user confirming (or rejecting) the predicted classification, 3. \textit{Human Moderation}: the discussion admins can choose to auto-validation mode (standard moderation for preventing abusive language) or complete manual, where each reporting log message is checked before published.

This hybrid approach has proven effective in the Municipio 9 pilot since February 2025. The system maintains human oversight while using NLP to streamline the categorisation and prioritisation process. 

\subsection{Smart Reporting with Customisable Widgets}

\begin{figure}[htbp]
    \centering
    \begin{minipage}{0.28\textwidth}
        \centering
        \includegraphics[width=\textwidth]{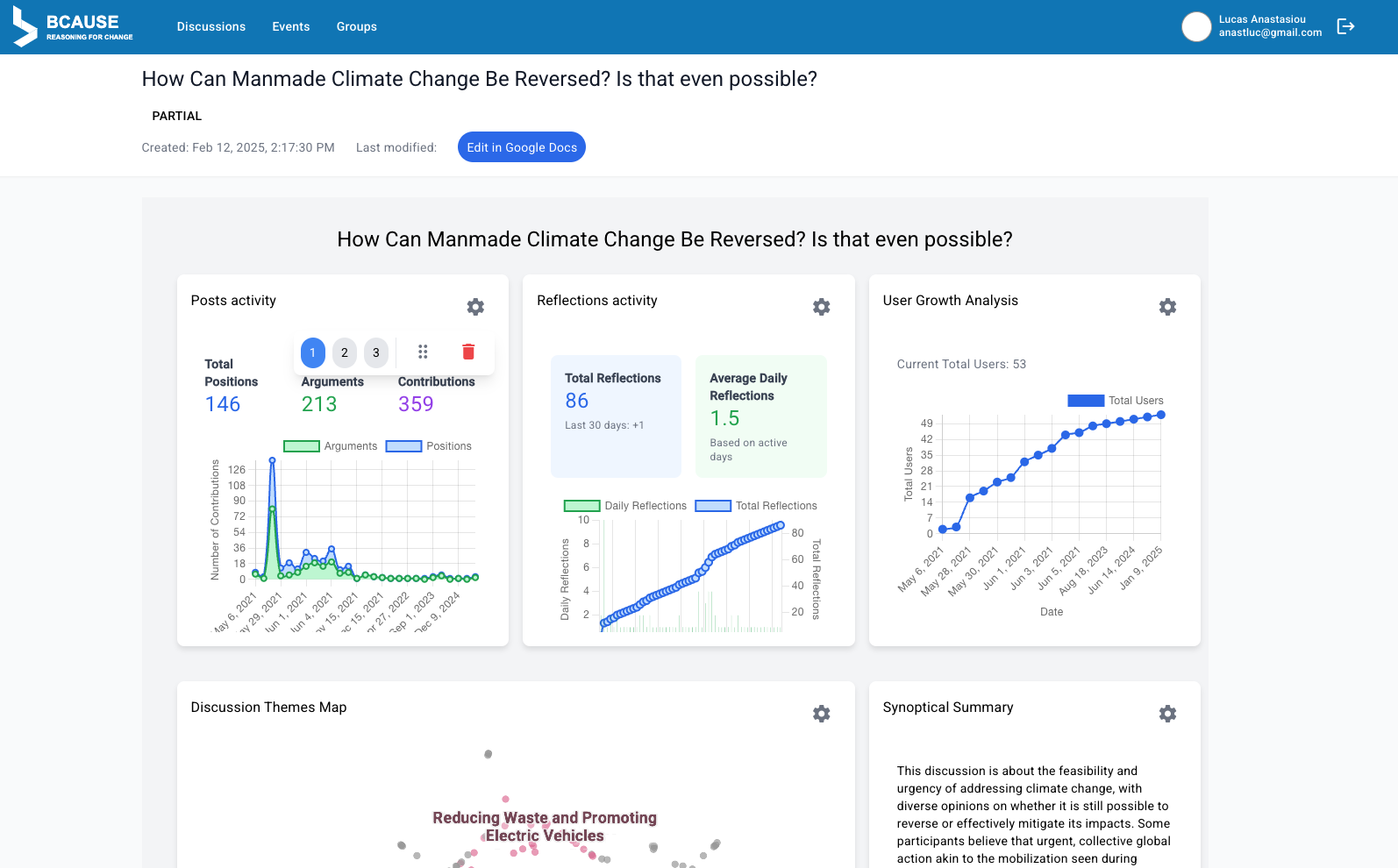}
    \end{minipage}
    \hfill
    \begin{minipage}{0.28\textwidth}
        \centering
        \includegraphics[width=\textwidth]{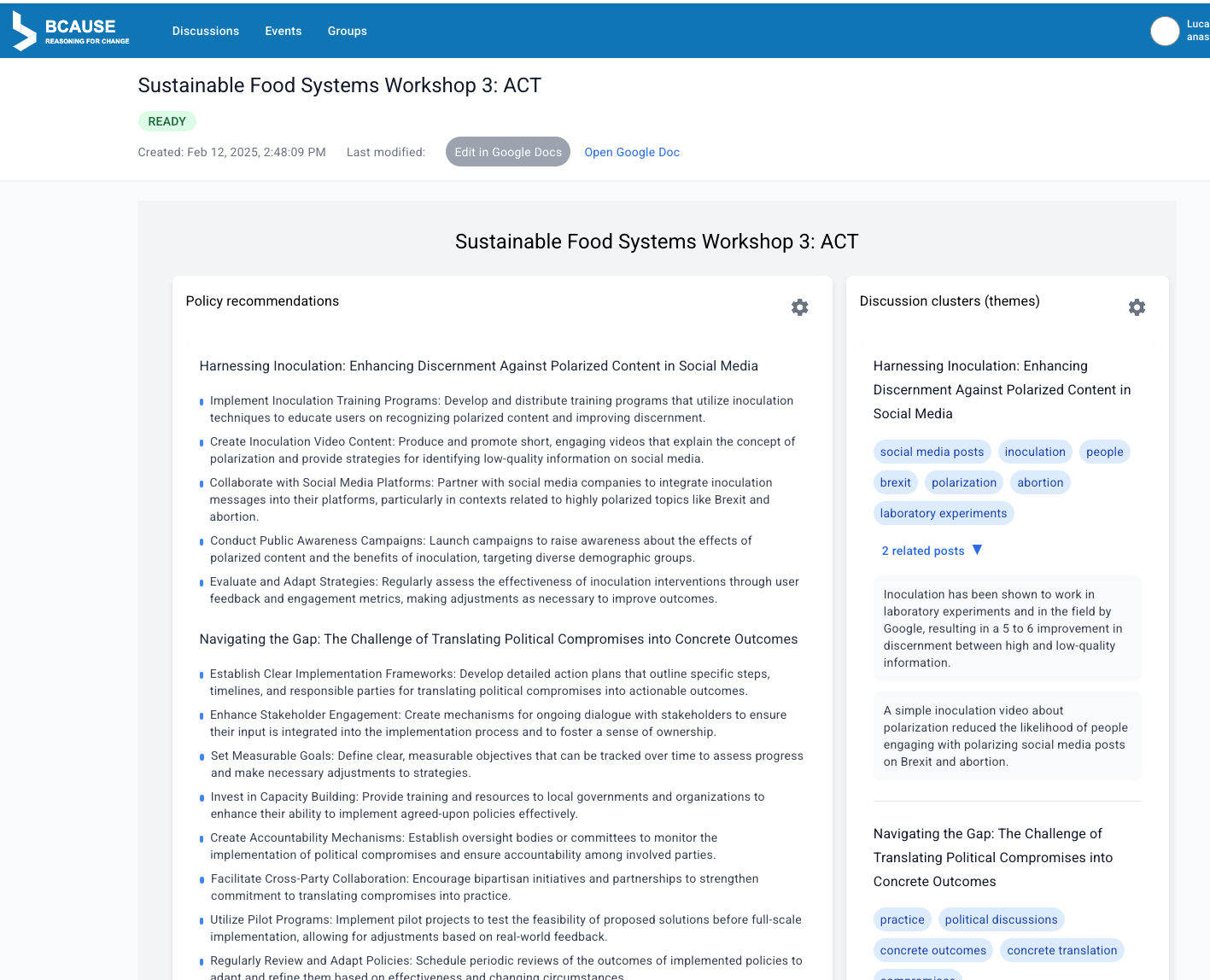}
    \end{minipage}
    \hfill
    \begin{minipage}{0.28\textwidth}
        \centering
        \includegraphics[width=\textwidth]{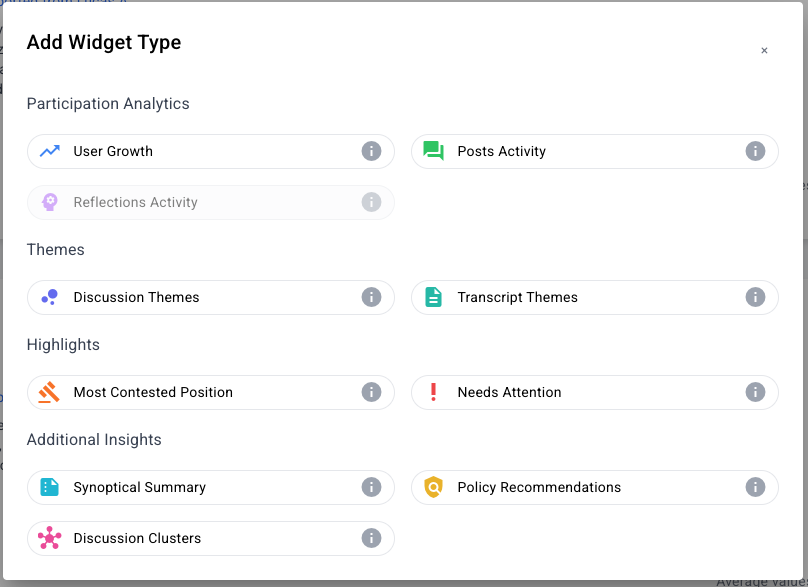}
    \end{minipage}
    \caption{Smart reporting interface featuring customizable widgets: (left) widgets with participation metrics and trend analysis, (centre) detailed discussion summary and argument clusters, and (right) available widgets selector menu}
    \label{fig:three-widgets}
\end{figure}

As a following step after the discussion stage in the overall deliberation, BCause implements an advanced reporting system that empowers moderators to generate comprehensive, interactive dashboards for analysing deliberative discussions. Automated reporting is a key enactor of human Sensemaking \cite{llinas2014survey}; however, there are various concerns in correctly incorporating it into deliberation processes\cite{anastasiou2021making}. The system features a modular approach with customisable widgets that can be tailored to specific analytical needs.

\subsection{Report Generation and Structure}
Moderators can initiate report generation at any point during a discussion, creating temporal snapshots that capture the state of deliberation at specific moments. Each report is structured as an interactive dashboard comprising various widgets that draw upon both raw discussion data and analyses.
The dashboard interface offers significant flexibility through (i)Drag-and-drop widget reorganisation, (ii) Resizable widgets to emphasise specific analytics, (iii) Individual widget export as .png/.pdf files, and (iv) complete dashboard export (and further edit and refine) in Google Docs. 
Beyond some analytical widgets like User growth over time, Posts and comments activity, engagement (reflections) progression and other participation analytics, there are content statistics widgets offering Agreement tracking (over each position posed in the debate), Position-argument distribution, and Position agreements distribution across discussion

However, the AI specific widgets are outlined as:
(i) \textit{Synopsis Widget}: Provides immediate context through a dynamic synoptical summary of the discussion, enabling quick understanding of scope and implications.
(ii) \textit{Discussion Themes Widget}: Presents theme analysis in two complementary views: Hierarchical tree structure showing thematic relationships and Detailed list format with theme-specific keywords and related posts
(iii) \textit{Points of Interest}: Highlights critical discussion elements as algorithmically identified: Most consensual point, Most opposed point and the Most contested point
(iv) \textit{Argument Network}: The combination of the discussion and the uploaded transcripts are ran through an argument mining pipeline and the main claims and premises are identified. Those are later visualised as an argument network using node-based representation of argumentative statements and colour-coded edges showing support/attack relationships

\section{Discussion}

We presented a series of three key genAI integrations into BCause discussion platform.
As a foundational design principle, BCause employs a hybrid approach where AI \textit{augments rather than replaces} human decision-making in deliberative processes.

In implementation, BCause's integrates seamlessly various genAI components exemplifying this human-AI partnership. While AI handles the initial processing of deliberative content, human moderators retain control over how arguments are structured and presented. 
In the reporting module for example, while AI performs the initial clustering and summarization of discussions, human moderators maintain final approval over visualizations and the affordance to edit the final report output ensures human verification before sharing with policy makers. 
Similarly, in the transcript import process, discussion moderators maintain quality control over AI predictions through a rigorous two-step validation protocol that ensures accuracy and reliability of imported content, while they firstly calibrate the AI prediction parameters to their needs.

This careful balance preserves the benefits of AI's processing capabilities while maintaining human judgment over the deliberation content.
This human-centred approach yields several benefits, particularly in building trust and ensuring democratic legitimacy. However, challenges persist, particularly in optimising the human-AI workflow (in terms of user interface, user experience and service design) in such way that it does not cognitively overload human moderators, it scales-up efficiently, while limiting inaccuracies. 
Future work will focus on rigorously evaluating these aspects, including comparative benchmarking against existing deliberation tools, assessing computational performance and scalability, and providing statistical validation of the AI-generated argument structures' accuracy and coherence.

\begin{acks}
This research was funded by the European Commission under the Horizon Europe Programme and by the UK Government’s Horizon Europe Guarantee scheme (Reference Number: 10048874) in the context of the ORBIS Project (GA: 101094765) on ``Augmenting participation, co-creation, trust and transparency in Deliberative Democracy at all scales''. The content of this publication does not necessarily reflect the views of the European Commission. Responsibility for the information and views expressed therein lies entirely with the authors.
\end{acks}

\bibliographystyle{ACM-Reference-Format}
\bibliography{biblio}

\appendix

\end{document}